\documentclass[conference]{IEEEtran}

\IEEEoverridecommandlockouts

\usepackage{amsmath}   
\usepackage{amssymb}
\usepackage{mathrsfs}

\usepackage{balance}   
\usepackage{flushend}

\usepackage{cite}      

\usepackage{ifpdf}     

\usepackage{multicol}
\usepackage{makeidx}
\usepackage{color}
\usepackage{pseudocode}
\usepackage{epsf}
\usepackage{subfigure}

\usepackage{comment}

\ifCLASSINFOpdf
   \usepackage[pdftex]{graphicx}
   \graphicspath{{../fig/pdf/}}
   \DeclareGraphicsExtensions{{.pdf}}
\else
   \usepackage[dvips]{graphicx}
   \graphicspath{{../fig/eps/}}
   \DeclareGraphicsExtensions{{.eps}}
\fi

\includecomment{nonDraftFigure} 
\excludecomment{draftFigure}

\makeatletter
\def\normalsize{\@setfontsize{\normalsize}{10bp}{10.00pt}}
\normalsize
\makeatother

\DeclareMathAlphabet{\mathpzc}{OT1}{pzc}{m}{it}

\newcommand{\be}{\begin{eqnarray}}
\newcommand{\ee}{\end{eqnarray}}

\begin{document}

\title{Reduced Complexity Detection for Network-Coded Slotted ALOHA using
Sphere Decoding
 \vspace{-0.15cm} }

\author{
Terry Ferrett and
Matthew C. Valenti,\\
West Virginia University, Morgantown, WV, USA.
\vspace{0cm}
}

\maketitle

\begin{abstract}
Network-coded slotted ALOHA (NCSA) is a refinement to the classic slotted ALOHA protocol which
improves throughput by enabling multiple source transmissions per ALOHA slot using
physical-layer network coding (PNC). 
The receiver detects the network-coded combination of bits during every slot and
recovers information bits by solving a system of linear equations.
This work develops a receiver capable of detecting the network-coded combination
of bits during a slot considering an arbitrary number of sources, orthogonal modulation, and a block fading channel.
Maximum-likelihood detection of the network-coded symbol at the receiver becomes
complex as the number of sources is increased.  To reduce this complexity, sphere
decoding is applied at the receiver to limit the number of constellation symbols the receiver must
consider for detection. The system is simulated for two modulation orders and two through five sources,
and error-rate performance results are provided.
\end{abstract}

\IEEEpeerreviewmaketitle


\section{Introduction}\label{sec:1}

Slotted ALOHA (SA) is a multiple-access protocol in which several sources
transmit information to a receiver over a timespan
denoted as a \emph{frame}.
The frame is divided into several discrete timespans denoted as \emph{slots}, and sources
align their packet transmission times to fall within the slots.
The protocol is used in a variety of applications, such as satellite communication.

Recently, several enhancements to classical slotted ALOHA have appeared 
in the literature.
Under the original formulation of SA reception of only one packet per slot is assumed,
and reception from two or more sources in a single slot is treated as interference and discarded.  
All of the enhancements are based on relaxing the assumption of
one packet reception per slot and allowing multiple, improving throughput.
In \cite{paolini:2015}, each source transmits its packet
in multiple slots, deliberately causing interference,
and the receiver subtracts non-interfered packets from
interfered ones to recover all source packets, a process
known as \emph{successive interference cancellation} (SIC).

A technique has been introduced which improves
the throughput of slotted ALOHA using physical-layer network coding (PNC) \cite{zhang2:2006},
denoted as \emph{network-coded slotted ALOHA} (NCSA) \cite{yang:2015}.
Intuitively, throughput is improved by allowing more than one
source to transmit a packet during a particular ALOHA slot.
For a particular ALOHA frame, the sum of packets in each slot is modeled
as a system of linear equations, and the receiver solves the system
to recover the packets transmitted by each source.
The NCSA algorithm is formulated
under the assumption that the detector at the receiver
can perform PNC on the packets received in each slot,
but does not describe a technique for doing so.  

Other works have considered the application of PNC to slotted ALOHA
PNC and sphere decoding, and modulation orders for PNC beyond two.
In \cite{cocco:2011}, two schemes are proposed combining PNC and slotted ALOHA,
one for terrestrial networks, and one for satellite networks, which require distinct
control information.
In \cite{mejri:2014}, a lattice-based PNC scheme utilizing compute-and-forward
is developed.
Higher-order modulations for a system combining PNC and multiuser detection
are considered by \cite{pan:2015}.

The main contribution of the current work
is to develop a receiver capable of performing the 
PNC operation, forming a \emph{network-coded packet} for each slot.
Our work considers a specific modulation, $M$-ary orthogonal modulation.
Previous work has thoroughly developed a PNC receiver for
frequency-shift keyed modulation under the case of relaying information between two users \cite{vtf:2011} \cite{ferrett:2013},
and this work extends the previous, considering an arbitrary number
of users.

Detecting the network-coded packet for a particular slot
requires performing detection on the sum of symbols transmitted
by each source.
The symbol determined by performing detection on this sum constellation
is denoted as the \emph{super-symbol}.
In a fading environment where the channels between each source and the destination
have independent fading coefficients, the number of possible super symbols is exponential
in the number of sources.
 
Maximum likelihood (ML) detection implemented by comparing
all possible super symbols to the received symbol 
clearly becomes intractable for increasing modulation order and number of sources.
This motivates the development of
a detection scheme with reduced complexity.
A well-known detection technique having complexity
which is independent of the constellation size is
\emph{sphere decoding} \cite{hochwald:2003}.
This work develops a sphere decoder which detects the super-symbols comprising
the network-coded packets in an ALOHA slot.
Detecting the transmitted super symbol enables recovery
of the bits in the network-coded packet.

The rest of the this work is presented as follows.
Section \ref{sec:2} describes the system model.
Section \ref{sec:3} develops the sphere decoder for physical-layer network
coding. Section \ref{sec:4} presents simulation results, and Section \ref{sec:5}
provides concluding remarks.

\section{System Model}\label{sec:2}
In this section, the modeling assumptions used throughout the work are developed.
The modulation scheme, channel model, and source and destination architectures are presented.
 
\begin{figure}[t]
  \begin{center}
    \includegraphics[width=6cm]{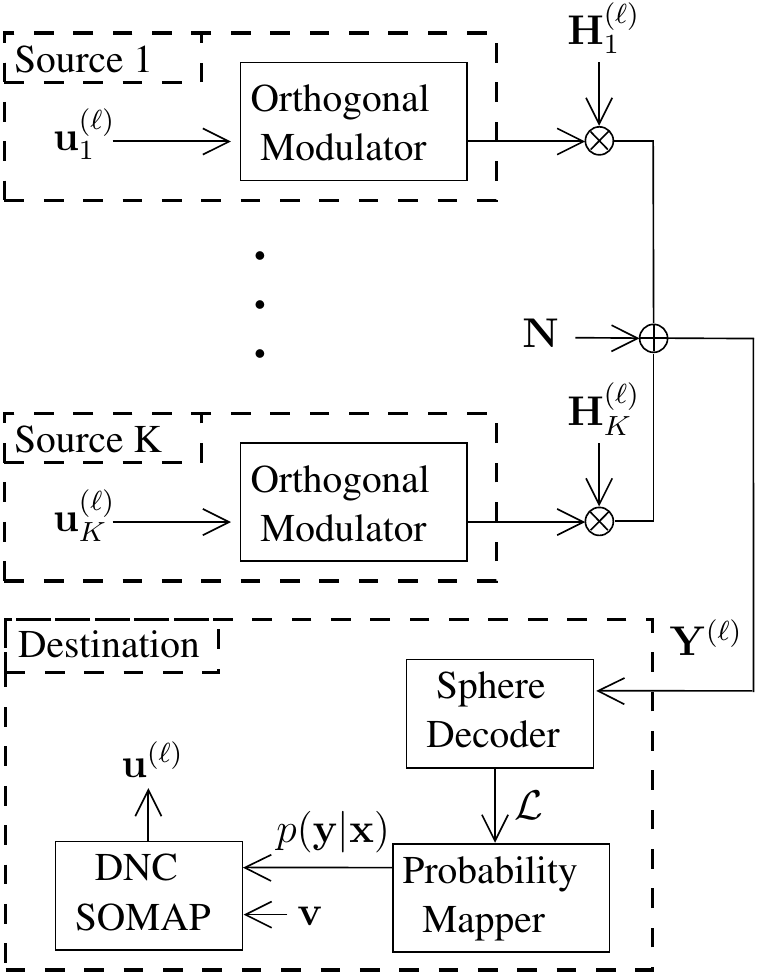}
  \end{center}
  \caption{System Model}
  \label{fig:sysm}
\end{figure}

\subsection{Transmission by Source Nodes}

The \emph{source} nodes $\mathcal{N}_k, \ k \in \{1,2,...,K\}$ generate length-$L$ binary information sequences $\mathbf{u}_k = [u_{k,1}, ..., u_{k,L}]$. 
Let $\mathcal{D} = \{ 0,...,M-1\}$ denote the set of integer indices corresponding to each orthogonal dimension, where $M$ is the modulation order.
The number of bits per symbol is $\mu = \log_2M$.
The information sequences $\mathbf{u}_k$ from each source are divided into $N_q = L/\mu$ sets of bits, which are passed to an $M$-ary orthogonal modulator.
The modulator maps each set of bits to an $M$-ary symbol $q_{k,i} \in \mathcal{D}$, where $k$ denotes the source, and $i$ denotes the $i$-th symbol, constructing a frame.

The symbols are represented in discrete time by the set of column vectors $\mathbf{x}_{k,i}$.
The column vector $\mathbf{x}_{k,i}$ is length $M$, contains a 1 at vector position $q_{k,i}$, and $0$ elsewhere.
The modulated frame from source $\mathcal{N}_k$ is represented by the matrix of symbols $\mathbf{X}_k =  [\mathbf{x}_{k,1}, ..., \mathbf{x}_{k,N_q}]$, having dimensionality $M \times N_q$.

\subsection{Channel Model}

All of the channels in the system are modeled as \emph{block-fading} channels.  A block is defined as a set of $N$ symbols
that all experience the same fading coefficient.
The duration of each block corresponds roughly to the channel coherence time.
The frame $\mathbf X_k$ transmitted by source $\mathcal N_k$ may be partitioned into $N_b=N_q/N$ blocks according to  
\begin{eqnarray}
  \mathbf{X}_{k}
  & = &
  \left[
    \begin{array}{ccc}
      \mathbf{X}_{k}^{(1)} & ... & \mathbf{X}_{k}^{(N_b)}
    \end{array}
  \right]
\end{eqnarray}
where each block $\mathbf{X}_{k}^{(\ell)}, 1 \leq \ell \leq N_b,$ is an $M \times N$ matrix, and the values
of $N_q$ and $N$ are chosen such that $N_b$ is an integer.
The vector of information bits mapped to the symbols in this block is denoted as $\mathbf{u}_k^{(\ell)}$, containing
$N \mu$ bits.
The channel associated with block $\mathbf{X}_{k}^{(\ell)}$ is represented by the $M \times M$ diagonal matrix 
$\mathbf{H}_{k}^{(\ell)} = \text{diag}( h_k^{(\ell)} )$, where $h_k^{(\ell)} = \alpha_k^{(\ell)} e^{j \theta_k^{(\ell)}}$, 
$\alpha_{k}^{(\ell)}$ is a real-valued fading amplitude distributed Rayleigh($1/\sqrt{2}$) and $\theta_k^{(\ell)}$ 
is a random phase shift distributed $U(0,2\pi]$.  
It is assumed that the frames transmitted by the sources are received in perfect synchronization
at the destination.
The $\ell^{th}$  block at the output of the relay receiver's matched-filters is then 
\begin{eqnarray}\label{eqn:RelayRecSignal}
  \mathbf{Y}^{(\ell)} 
  & = &  \sum_{k=1}^K  \mathbf{H}_{k}^{(\ell)} \mathbf{X}_{k}^{(\ell)} + \mathbf{N}^{(\ell)}
\end{eqnarray}
where $\mathbf N^{(\ell)}$ is an $M \times N$ noise matrix having elements which are i.i.d. circularly-symmetric complex Gaussian random variates.

\vspace{2cm}

\subsection{Destination Reception}

The goal of the destination receiver is to detect the network-coded sum of bits transmitted by the sources

\begin{align} \label{eq:nc_bit_sum}
  \mathbf{u} &= \sum_{k=1}^K \mathbf{u}_k
\end{align}

\noindent where the sum is taken modulo-2.
The elements of vector $\mathbf{u}$ are then

\begin{align} \label{eq:nc_bit_sum_one_elem}
  u_i = u_{1,i} \oplus u_{2,i}\oplus ... \oplus u_{K,i} \ \ i \in \{1, ..., L\}
\end{align}

\noindent where $\oplus$ is the exclusive-or operation. 
The network-coded bits experiencing the $\ell$-th fading block
are denoted as $\mathbf{u}^{(\ell)}$, containing $N \mu$ bits.

Consider a single symbol transmission period.
The constellation formed by the sum of all possible combinations of symbols 
which can be transmitted by the sources is denoted as the \emph{super-symbol constellation} and is defined as

\begin{align}
  \mathbf{x} = \sum_{k=1}^K h_k \mathbf{x}_k.
\end{align}

\noindent Under orthogonal modulation, the cardinality of this constellation is $M^K$.
The network-coded bits represented by each super-symbol are defined as

\begin{align}
  b_i(\mathbf{x}) = b_i(\mathbf{x}_1) \oplus b_i(\mathbf{x}_2) \oplus ... \oplus b_i(\mathbf{x}_K)
\end{align}

\noindent where the mapping $b_i(\cdot)$ selects the $i$-th bit mapped to the symbol,
$b_i(\mathbf{x}) = u_i$, and $b_i(\mathbf{x}_k) = u_{k,i}$.

The receiver performs detection on a frame of channel observations $\mathbf{Y}$ one observation at
a time.
A single observation is represented by a single column of $\mathbf{Y}$ and is denoted as $\mathbf{y}$.
The operations performed on every channel observation are the same, thus, dependence on a particular
symbol interval is dropped.
A channel observation is comprised of the sum of symbols $q_k$ transmitted by the sources during
a single symbol interval.

A conventional demodulator performs detection by computing the conditional probability of receiving every possible super-symbol 
$p(\mathbf{y}|\mathbf{x})$, an exhaustive computation which grows exponentially in the number of sources \cite{ferrett:2013}.
Sphere decoding reduces the number of required computations by determining a \emph{list} $\mathcal{L}$ of
$N_S$ candidate super-symbols which fall within a specified \emph{radius} from
the received channel observation.
Details of the sphere decoder are provided in Section \ref{sec:3}.

The symbol probabilities are transformed to the set of $\mu$ \emph{log-likelihood ratios} (LLRs) associated with each network-coded bit mapped to a particular super-symbol.
Denote this operation as \emph{digital network-coded soft mapping} (DNC-SOMAP) \cite{ferrett:2013}.
To detect the network-coded bits, a hard decision is made on each LLR.
The LLR $z_i$ of the $i$-th network-coded bit mapped to the super-symbol is computed as

\vspace{-2mm}
\begin{align} \label{eq:somap_out}
  z_i & =  \underset{\begin{subarray} (\mathbf{x}: u_i = 1 \end{subarray}}{\operatorname{max}  \hspace{-0.5mm} *} \left[ \log p(\mathbf{y} | \mathbf{x}) + \sum_{\begin{subarray} jj=0 \\ j \neq i\end{subarray}}^{\mu-1} u_j v_j\right] \nonumber \\ & -\underset{\begin{subarray} (\mathbf{x}: u_i = 0 \end{subarray}}{\operatorname{max}  \hspace{-0.5mm} *} \left[ \log p(\mathbf{y} | \mathbf{x}) + \sum_{\begin{subarray} jj=0 \\ j \neq i\end{subarray}}^{\mu-1} u_j v_j  \right].
\end{align}

\noindent where $p(\mathbf{y} | \mathbf{x}) \in \mathcal{L}$, $u_i$ and $u_j$ are the $i$-th and $j$-th network-coded bits mapped to super-symbol $\mathbf{x}$, and $v_j$ is the a-priori LLR of network-coded bit $u_j$ which may be fed back from soft-output channel decoder. 
This work considers hard-decision decoding, so all $v_j = 0$.
\noindent The \emph{max-star} operator is defined as

\vspace{-2mm}
\begin{align}
  \underset{i}{\operatorname{max} \hspace{-0.5mm}*} \{ x_i \} = \log \left\{ \sum_i e^{ x_i } \right\}
\end{align}

\noindent where the binary max-star operator is  $\max*(x,y) = \max(x,y) + \log( 1 + e^{ -|x-y| } ) $ and
multiple arguments imply a recursive relationship; for example: $\max*(x,y,z) = \max*( x, \max*(y,z) )$.

The value taken by the pdf $p(\mathbf{y}|\mathbf{x})$ is dependent on the available channel state information.
In this work it is assumed that the destination has perfect knowledge of the channel gains and noise variance

\begin{align}\label{eq:ch_pdf}
p(\mathbf{y} | \mathbf{x}) = \left( \frac{1}{\pi N_0}\right)^M \exp\left\{ -\frac{1}{N_0} ||\mathbf{y} - \mathbf{x} ||^2 \right\} 
\end{align}

\section{List Sphere Decoder}\label{sec:3}

  The goal of the list sphere decoder (LSD) \cite{hochwald:2003} is to determine the set of super-symbols $\mathcal{L}$ which fall within a specified distance $r$ from the channel observation.  
  Specifically, the LSD searches for super-symbols satisfying the following relation

  \begin{align}\label{eq:vec_dist}
    (\mathbf{x} - \mathbf{y})^* (\mathbf{x} - \mathbf{y}) \leq r^2
  \end{align}

  \noindent where $(\cdot)^*$ denotes the conjugate transpose operation, and $r$ is the \emph{radius} from the received channel observation $\mathbf{y}$.
  Knowledge of the fading coefficients between each source and the destination is used to construct the set
  of constellation points which fall within the specified radius.

  The efficiency of the sphere decoder algorithm is based on recursively computing Eq. (\ref{eq:vec_dist}) as
  the summation 
  
  \begin{align}\label{eq:sd_terms}
    \sum_{k=1}^{M} | x_k - y_k + \sum_{j = k + 1}^M x_j - y_j|^2 \leq r^2
  \end{align}
  
  where the subscripts $k$ and $j$ denote the $k$-th and $j$-th components of vectors $\mathbf{x}$ and $\mathbf{y}$.
  The term $ k = M $ yields

  \begin{align} \label{eq:i_M}
    |x_M - y_M | \leq r
  \end{align}

  To efficiently compute whether a super-symbol in the received constellation falls within the specified
  radius $r$ from the channel observation, a parametric description of the constellation is required.
  Recall that the modulation scheme is modeled using an $M$-dimensional complex vector.  Since each vector
  component is complex valued, each may be regarded as a two-dimensional vector space having 
  in-phase and quadrature components.  The LSD recursively detects the set of in-phase and quadrature components
  within each vector component  which fall within $r$ to form the list of received points. 

  Note that the $m$-th component of $\mathbf{x}$ takes the following value
  in the absence of noise

  \begin{align} \label{eq:rec_fc}
    x_m = \sum_{k=1}^{K} h_k x_{k,m}
  \end{align}
 
  \noindent where $x_{k,m}$ is the $m$-th component of the symbol transmitted by source $\mathcal{N}_k$. 
  Recall that $x_{k,m}$ takes value $1$ if the $k$-th source transmitted in the $m$-th
  dimension and $0$ otherwise.  It follows that $x_m$ will take one of $2^K$ values during a single symbol period, drawn
  from the sums of all possible subsets of $\mathbf{h} = \{ h_1, h_2, ..., h_K \}$.

  Consider the $m$-th vector component formed from the sum of a particular subset of $\mathbf{h}$.  
  The component may be expressed as $x_{m}^{(\ell)} = d_{\ell} e^{i \theta_{\ell}}$,
  where $\ell$ denotes a subset of $\mathbf{h}$.
  Denote the $m$-component of $\mathbf{y}$ as $y_{m} = \hat{d}_m e^{i \hat{\theta}_{m}}$.
  Then Eq. (\ref{eq:i_M}) may be expressed as
 
  \begin{align}
    |x_{M}^{(\ell)} - y_M |^2 = d_{\ell}^2 + \hat{d}_M^2 -  2 d_{\ell} \hat{d}_M \cos( \theta_{\ell} - \hat{\theta}_M ) \leq r^2
  \end{align}

  \noindent rearranging 

  \begin{align}\label{eq:bnd_i_M}
    \cos( \theta_{\ell} - \hat{\theta}_M ) \geq \frac{ d_{\ell}^2 + \hat{d}^2_M - r^2 }{ 2 d_{\ell} \hat{d}_M }
  \end{align}
  
  \noindent Define the right-hand side of Eq. (\ref{eq:bnd_i_M}) as

  \begin{align}
    \eta = \frac{ d_{\ell}^2 + \hat{d}^2_M - r^2 }{ 2 d_{\ell} \hat{d}_M }
  \end{align}

  \noindent The value of $\eta$ determines whether constellation component $x_{M}^{(\ell)}$ falls within
  the radius as

  \begin{align}
     \begin{cases}
       \eta > 1 & \text{Not within radius}  \\
       -1 \leq \eta \leq 1 &  \text{If } \hat{\theta}_M - \cos^{-1} \eta  \leq \theta_{\ell} \leq \hat{\theta}_M + \cos^{-1} \eta  \\
       \eta < -1 & \text{Within radius}.
    \end{cases}
  \end{align}

\begin{figure}[t]
    \begin{center}
      \includegraphics[width=6cm]{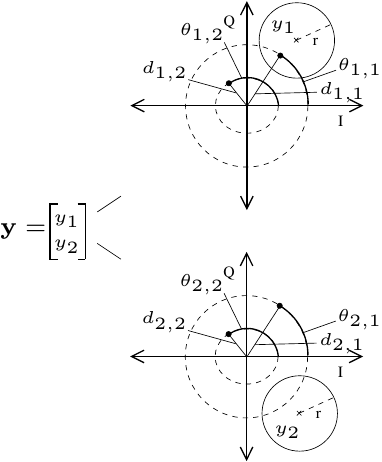}
    \end{center}
    \caption{Sphere Decoding Example: M=2}
    \label{fig:sd}
  \end{figure}

  Thus, the super-symbol is not within the radius when $\eta$ is greater than $1$,
  always within the radius when $\eta$ is less than $-1$, and conditionally within the radius
  when $\eta$ falls between $-1$ and $1$.  An example is shown in Fig. \ref{fig:sd}.

  \vspace{2cm}
  For terms $k < M$ Eq. (\ref{eq:sd_terms}) becomes
  
  \begin{align}
    |x_{k}^{(\ell)} - (y_k - a_j)| \leq \sqrt{r^2 - b_j}
  \end{align}

  where the terms $a_j$ and $b_j$ are

  \begin{align}
    a_j &= \sum_{j=k+1}^M x_{j}^{(\ell)} - y_j\\
    b_j &= \sum_{j=k+1}^M |x_{j}^{(\ell)} - y_j + \sum_{o=j+1}^M x_{o}^{(\ell)} - y_m|^2
  \end{align}
  
  \noindent where $x_{j}^{(\ell)}$ is the constellation component selected at the previous $j$-th dimension.
  The terms $a_j$ and $b_j$ can be efficiently accumulated during computation of every $k$-th term
  of (\ref{eq:sd_terms}).

  The algorithm proceeds as follows. Beginning with $k=M$, the constellation components $x_{k}^{(\ell)}$
  which fall with the radius $r$ are computed for all $\ell$, and one is selected.
  The value of $k$ is decremented by $1$, constellation components are computed for the new $k$, and a single
  component is selected once again.  If a $k$ value is reached where no points are found, the algorithm backtracks by adding
  $1$ to $k$ and selecting the next available constellation component. 

  Once $k=1$ is reached, a super-symbol is selected and its distance from the received point is computed. 
  If the list is not full, the point is added to the list.  If the list is full, the distance for the new point
  is compared to the longest in the list.  If the new super-symbol distance is shorter than the last in the list, the new
  super-symbol is added to the list, replacing the previous longest point.
  
  The algorithm terminates when distances are computed for all super-symbols falling within
  the radius.



\section{Simulation Results}\label{sec:4}

\begin{figure}[t]
  \begin{center}
    \includegraphics[width=8.8cm]{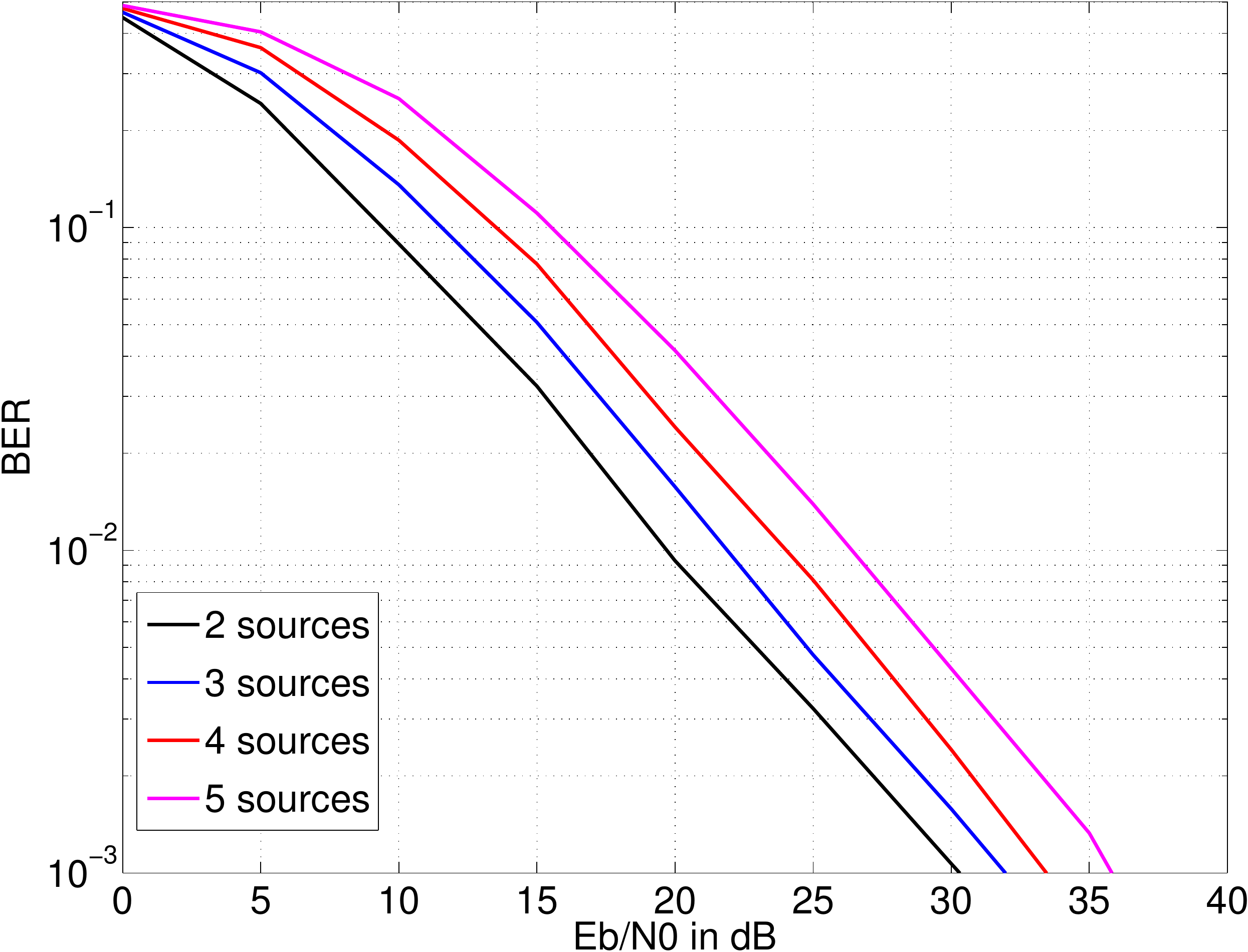}
  \end{center}
  \caption{Simulated error-rate performance for modulation order $M=2$.
    The number of sources considered is $K=\{2,3,4,5\}$.
    The information sequence length is $L=2304$.
    List sphere decoding uses $N_S = 5$ symbols per list.
    A sphere decoding radius $r = 4 N_0$ is utilized.}
  \label{fig:ber1}
\end{figure}

This section presents simulation results.
The system described in Section \ref{sec:2} is simulated
via the Monte Carlo method for several values of modulation
order and number of sources.
All signal-to-noise ratio values are expressed in terms of
energy-per-bit ($E_b/N_0$).

The sources are simulated using the following parameters.
The number of sources considered is $K=\{ 2, 3, 4, 5 \}$.
The information sequence length is $L = 2304$.
The orthogonal modulation orders are $M=\{ 2, 4 \}$, having $\mu = \{ 1, 2\}$
bits per symbol, respectively.
The number of symbols per frame is $N_q = 2304$ under $M=2$ and
$N_q = 1152$ under $M=4$.
During a single frame transmission, a uniformly random information
sequence $\mathbf{u}_k$ is generated for each source and modulated
to produce frames $\mathbf{X}_k$.

The frames are passed through the channel and corrupted.
The block size considered throughout is $N=2304$,
which was selected to reduce the required simulation time,
as every block requires exhaustively computing the values
of the fading coefficient described by Eq. (\ref{eq:rec_fc}).
The frames are corrupted by the fading coefficients, added,
and corrupted by noise according to Eq. (\ref{eqn:RelayRecSignal}).

\begin{figure}[t]
  \begin{center}
    \includegraphics[width=8.8cm]{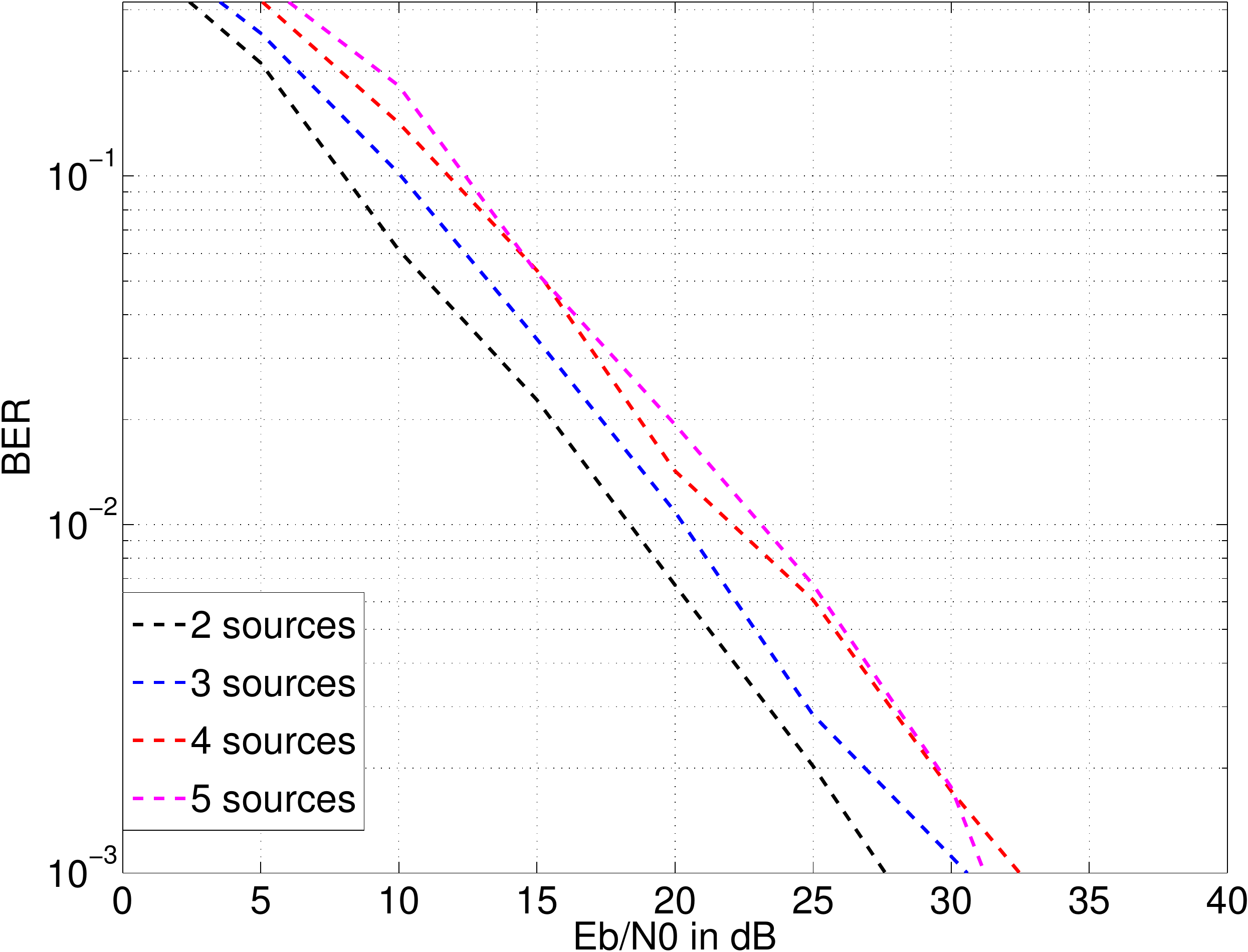}
  \end{center}
  \caption{Simulated error-rate performance for modulation order $M=4$.
  See the Fig \ref{fig:ber1} caption or Section \ref{sec:4} for simulation parameters.}
  \label{fig:ber2}
\end{figure}

For every symbol period in the frame, the sphere decoding algorithm
is applied to limit the number of super-symbols required for
computation by the DNC-SOMAP.
The size of the candidate list $\mathcal{L}$ is set to $N_S = 5$.
The sphere radius is set in proportion to the noise variance
$r = 2 B N_0$, where $B$ is a constant.  It was discovered
through experimentation that $B=2$ enabled successful
decoding in all cases.
The sphere decoding algorithm described in Section \ref{sec:3} is applied
to all symbol periods in the received frame, and the value $p(\mathbf{y}|\mathbf{x})$
is computed for every candidate super-symbol in all $\mathcal{L}$ according to Eq. (\ref{eq:ch_pdf}).

The values of $p(\mathbf{y}|\mathbf{x})$ are passed to the DNC-SOMAP to compute
the LLRs of the network-coded bits for the frame using the relationship
described by Eq. (\ref{eq:somap_out}).
The network-coded bits comprising a single frame $\mathbf{u}$ are computed
by making a hard decision on the LLRs.
The a-priori LLRs to the DNC-SOMAP $v_j$ are set to $0$ for all
simulations.

Figs. \ref{fig:ber1} and \ref{fig:ber2} show error-rate performance for modulation
orders $M=2$ and $M=4$, respectively.  
Considering $M=2$, performance degrades by approximately $1\ dB$ for every source
added to the system at error rate $10^{-3}$.
Case $M=4$ appears to exhibit similar behavior, as the difference in performance
between the best and worst performing curves are similar to the $M=2$ case.
The curves in the $M=4$ case are not as smooth as $M=2$, suggesting that more Monte Carlo 
trials are required to reduce variance.

\section{Conclusion}\label{sec:5}

This work has developed and demonstrated a sphere decoder formulation for detecting the network-coded
combination of bits transmitted by an arbitrary number of sources. The system has been developed
assuming orthogonal modulation and a block fading channel.  Simulation results demonstrate
that the error-rate performance of the system degrades in proportion to the number of sources.

Several open issues remain.  Sphere decoding in the considered system requires
calculating all possible combinations of fading coefficients, which requires
$2^K$ computations.  A more efficient approach is necessary for the sphere decoder
formulation to be practically efficient.  The value of the sphere decoding radius was
set in proportion to the channel noise variance, however, more efficiency may be obtained
by reducing the radius size.

\balance

\bibliographystyle{IEEEtran}
\bibliography{bibliography}

\end{document}